\documentclass[final,5p,times,twocolumn]{elsarticle}

\usepackage[utf8]{inputenc}
\usepackage{graphicx}
\usepackage{dcolumn}
\usepackage{bm}%
\usepackage{xcolor}

\begin{document}



\title{Electronic, optical, vibrational and thermodynamic properties of phaBN structure: a first principles study}

\author[UFRPE]{J. M. Pontes}
\author[UFCG1,UFCG2]{N. F. Fraz\~{a}o}
\author[UNB1,UNB2]{David L. Azevedo}
\author[UFRPE,KIT]{Jonas R. F. Lima}
\ead{jonas.lima@ufrpe.br}

\address[UFRPE]{Departamento de F\'{\i}sica, Universidade Federal Rural de Pernambuco, 52171-900, Recife, PE, Brazil}
\address[UFCG1]{Centro de Educa\c{c}\~{a}o e Sa\'{u}de, Universidade Federal de Campina Grande, 581750-000, Cuit\'{e}, PB, Brazil}
\address[UFCG2]{Unidade Acad\^{e}mica de F\'{\i}sica, Universidade Federal de Campina Grande, 58429-900, Campina Grande, PB, Brazil}
\address[UNB1]{Faculdade UnB Planaltina, Universidade de Bras\'{i}lia, 73345-10, Bras\'{i}lia, DF, Brazil}
\address[UNB2]{Instituto de F\'{\i}sica, Universidade de Bras\'{i}lia, 70919-970, Bras\'{i}lia, DF,  Brazil}
\address[KIT]{Institute of Nanotechnology, Karlsruhe Institute of Technology, D-76021 Karlsruhe, Germany}

\date{\today}


\begin{abstract}
In 2015, a new two dimensional (2D) carbon allotrope, called phagraphene, was theoretically proposed. Based on this structure, we propose here a new boron nitride structure called phaBN. It is composed by three types of rings: pentagons, hexagons and heptagons. We investigate the electronic, optical, vibrational and thermodynamic properties of phaBN using first-principles calculations in a density functional theory (DFT) framework. Our calculations revealed that the phaBN has an energy gap of 2.739 eV, which is almost half of the energy gap of the hexagonal boron nitride (h-BN), thus being a semiconductor material. By means of the optical, vibrational and thermodynamic properties, it was possible to observe the absorption interval, the stability of the structure and its formation process, respectively.
\end{abstract}

\maketitle


\section{Introduction}

The first successful experimental realization of graphene in 2004 \cite{Novoselov} open a new field of research, the two-dimensional (2D) materials. Since then, the number of 2D materials proposed and synthesized increases every year \cite{Naguib,Xu2018}. These materials present a variety of novel properties and phenomena and have been used in a plenty of applications. Looking to the electronic properties, 2D materials can be, for instance, metallic \cite{C7QM00548B}, semiconductor \cite{CHOI2017116,doi:10.1116/1.4982736,Manzeli2017,kolobov_2018}, insulator \cite{Illarionov2020}, semimetal \cite{20}, half-metal \cite{C6TC04490E,doi:10.1021/acs.nanolett.7b01367,PhysRevMaterials.3.084201} and superconductor \cite{Saito2016}.

The interest in 2D materials of nitrogen (N), boron (B) and carbon (C) compounds has increased due to the great potential that these materials have for the development of new nanotechnologies \cite{C7RA00260B,doi:10.1021/acsnano.6b08136}. The similarity between boron nitride structures and carbon allotropes is indisputable, with respect to its molecular geometries, strength, malleability and some other characteristics. This led to the realization of a boron nitride structure similar to graphene in 2005 \cite{Novoselov10451}, which is called hexagonal boron nitride (h-BN). Despite the structural similarity, the main difference between these structures is their electronic properties, since graphene is a semimetal with no energy gap and h-BN is an insulator with a gap of 5.955~eV \cite{Cassabois2016}. 

In 2015, a new 2D carbon allotrope was theoretically proposed using computer simulation, which was called phagraphene \cite{23}. The structure of phagraphene is composed of three different carbon rings: pentagons, hexagons and heptagons. This flat carbon allotrope is energetically comparable to graphene and more favorable than other carbon allotropes proposed in previous works due to its hybridization in $sp^2$. Since its proposal, phagraphene attracted great deal of attention, which can be confirmed by the number of works addressed to this subject \cite{C6CP04595B,C6RA05082D,doi:10.1021/acsami.7b04170,C6CP08621G,BAGHERI20188,LUO2017277,YUAN2017228,Podlivaev2016,LIU2016279,DONG2018206,RAJKAMAL2018775,Yuan_2018,doi:10.1021/acs.jpcc.6b05593,doi:10.1063/1.5029845,ROUHANI2020113710,LIU201730,SHEKAARI2020113979}. 

Using the structure of phagraphene as a model, we propose here a new material, the phaBN. This 2D structure has small undulations on its surface and is formed by boron and nitrogen atoms that bind only by simple bonds in which the number of atoms of the two elements are equal. Like the phagraphene, the phaBN also presents 5-6-7 rings in its structure. We investigated the electronic, optical and thermodynamic properties using first principles calculations. We obtained that the inclusion of the pentagons and heptagons in the structure of the h-BN to form the phaBN leads to the introduction of new electronic states inside of the energy gap, reducing the energy gap to 2.74~eV, which makes the phaBN a semiconductor. The phaBN brings possible optical and electronic applications.

This work is organized as follows: in Sec. 2, we present all the computational procedures used to obtain the properties of the systems. In Sec. 3, we show and discuss the results, which includes the electronic, optical, vibrational and thermodynamic properties of phaBN. The paper is summarized and concluded in Sec. 4.

\section{Method and Structure}

\begin{figure}
	\centering
	\includegraphics[width=\linewidth]{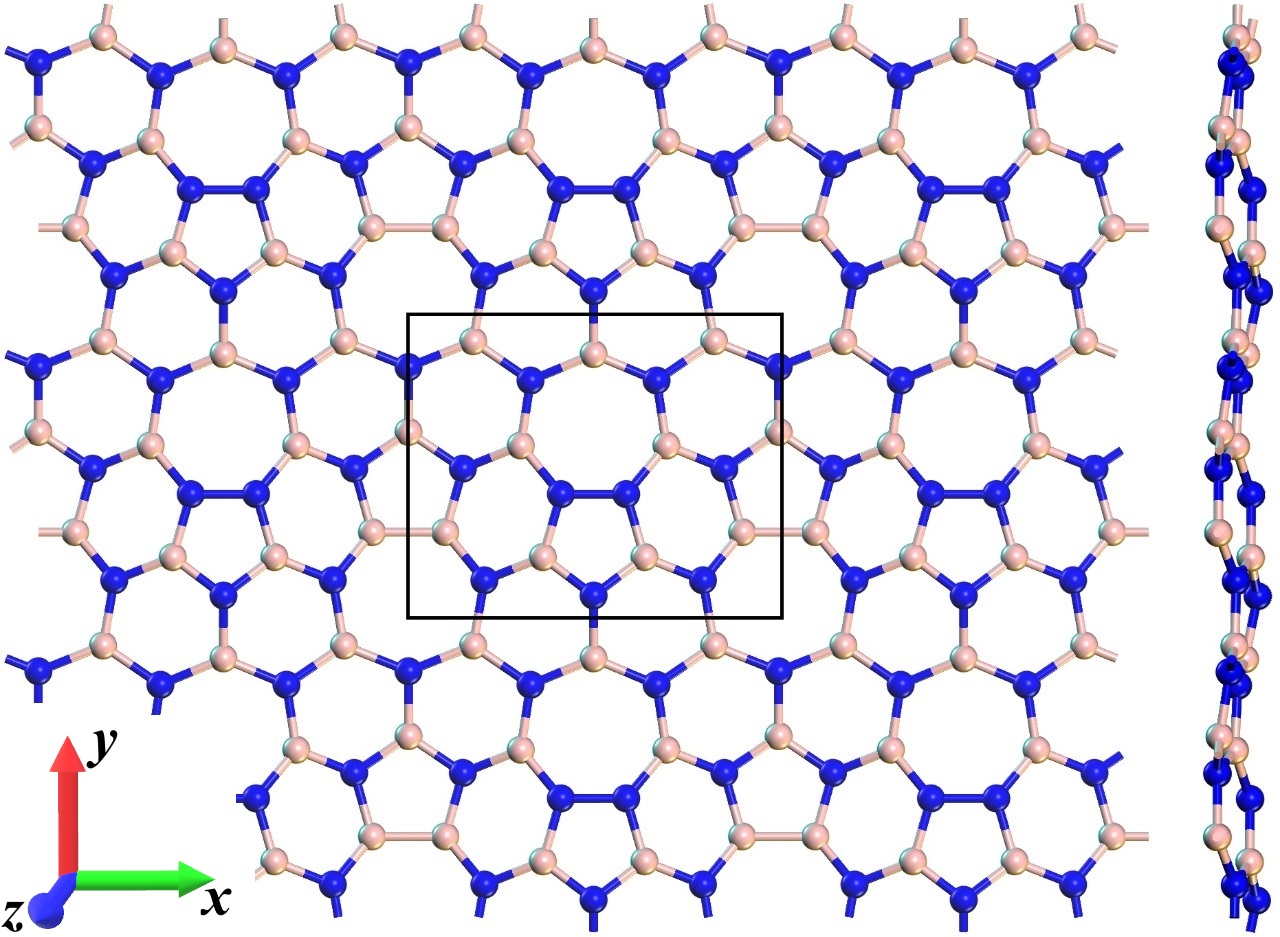}
	\caption{The crystalline structure of the phaBN. The blue (beige) circles represent the nitrogen (boron) atoms. The rectangular frame indicates the unit cell. In the right side we can see a lateral view of the phaBN.}
	\label{fig:phabn}
\end{figure}

In the present work, we investigate the phaBN structure, which is represented in Fig. \ref{fig:phabn}. The initial network parameters of the triclinic unit cell are given by: a = 8.163 \AA, b = 6.715 \AA~and c = 14.613 \AA, belonging to the space group P1. In order to perform the calculations, the CASTEP module of Material Studio suite was used, which is based on the DFT formalism \cite{33,34}. We choose the well-established Generalized Gradient Approximation (GGA) with the exchange-correlation functional of Perdew-Burke-Ernzerhof (PBE)\cite{35,42,43}. We also take into account Norm-conserving pseudopotential, usually harder to calculate, but it gives us better results than the Vanderbilt ultrasonic pseudopotentials and also allows calculations based on linear responses of phonon properties and polarizabilities \cite{70,72}. A cut-off energy limit of 770 eV was adopted for the set of plane wave bases to represent the Kohn-Sham orbitals. A sampling of K-points was done using a 14x14x1 Monkhorst-Pack grid for the evaluation of all integrals of the reciprocal space. The chosen grid had a very satisfactory result in the convergence of the electronic structure.

The optimization of the network parameters and the atomic positions were done seeking the lowest total energy for the unit cell of the phaBN. The optimization of the geometry was calculated according to the following limits of convergence for successive self-consistent steps: 0.001 eV/atom for the total energy variation, 0.1 eV/\AA~for maximum strength, 0.2 GPa for pressure and finally 0.005 for maximum atomic displacement. We also used the Broyden Fletcher-Goldfarb-Shanon (BFGS) minimizer for the calculation of unit cell optimization \cite{53}.

Once the unit cell is optimized, the analysis of the electronic structure of Kohn-Sham are done by means of the GGA-PBE approximation, following the recommendations of Refs. \cite{51,54}. We also obtain the dielectric function as well as optical absorption of light polarized in specific directions and in all directions (polycrystalline). Phonon and thermodynamic analyzes were used to examine the dynamic and thermal stability, respectively.

\section{Results}

\subsection{Electronic properties}

\begin{figure}
	\centering
	\includegraphics[width=\linewidth]{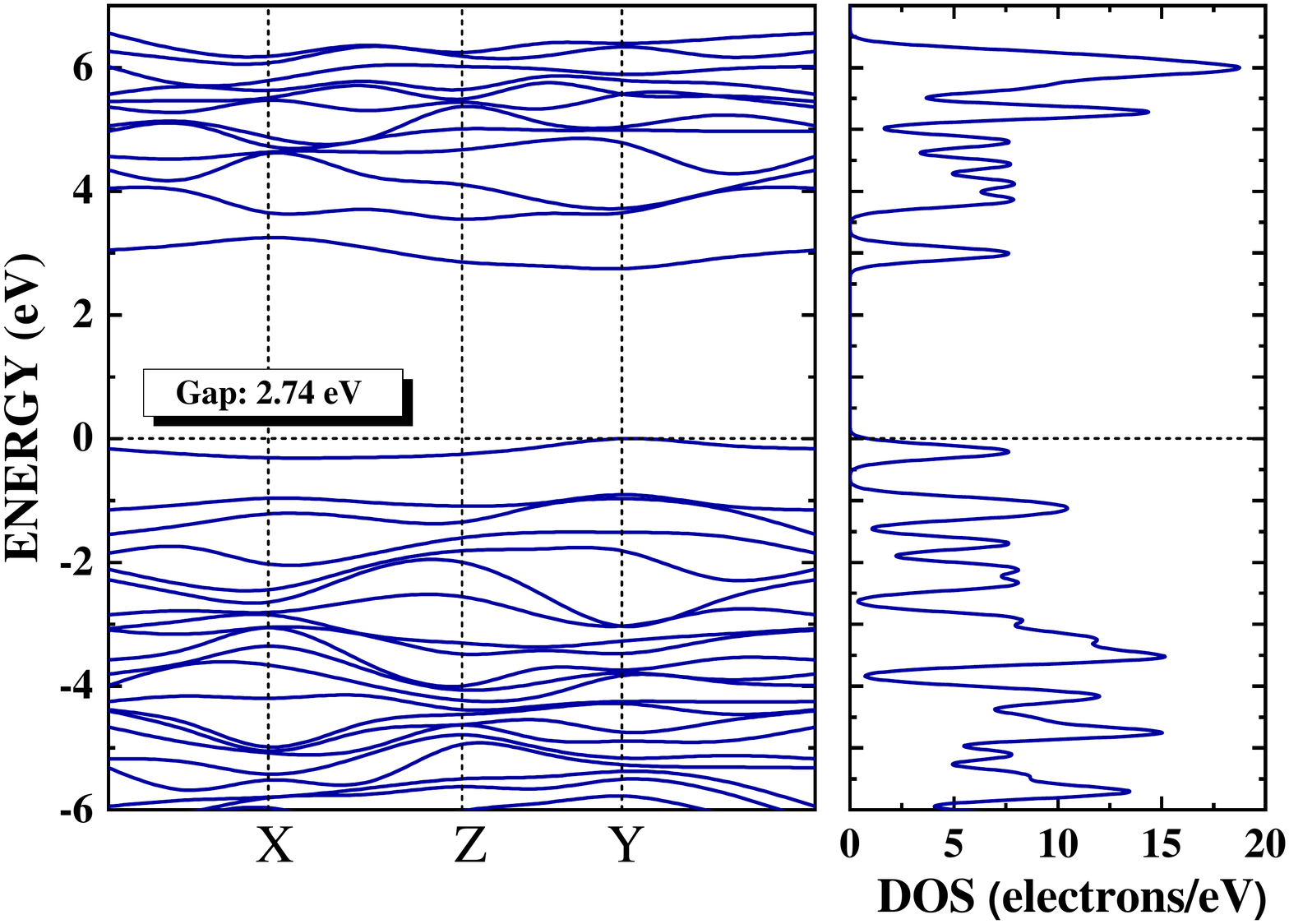}
	\caption{Representation of Kohn-Shan band structure and density of states of phaBN calculated using the GGA-PBE exchange correlation functional. The phaBN is a semiconductor with an energy gap of 2.74 eV.}
	\label{fig:ed}
\end{figure}

Let us first consider the electronic properties. In Fig. \ref{fig:ed} we have the band structure and the density of states (DoS) for the phaBN structure. The dashed horizontal line represents the Fermi level, which was shifted to zero. The paths used in the first Brillouin zone (BZ) were chosen using straight segments linking a set of high symmetric points. The chosen points are given by: $\Gamma$ (0,000, 0,000, 0,000), X (0,500, 0,000, 0,000), Z (0,500, 0,500, 0,000), Y (0,000, 0,000, 5,000).

We can see that the phaBN is a semiconductor with an energy gap of 2.74 eV. If we remove the highest occupied and the first unoccupied bands, the electronic structure of the phaBN becomes very similar to the h-BN. This indicates that the pentagonal and heptagonal pair of defects in the structure of the phaBN lead to the appearance of new electronic states inside of the energy gap of the h-BN. Such electronic states usually appear in the presence of impurities, which are introduced, for instance, by doping. In the phaBN, the pair of defects are inducing these states, since there is no doping and the number of N and B atoms are the same in the system. 

\begin{figure}
	\centering
	\includegraphics[width=\linewidth]{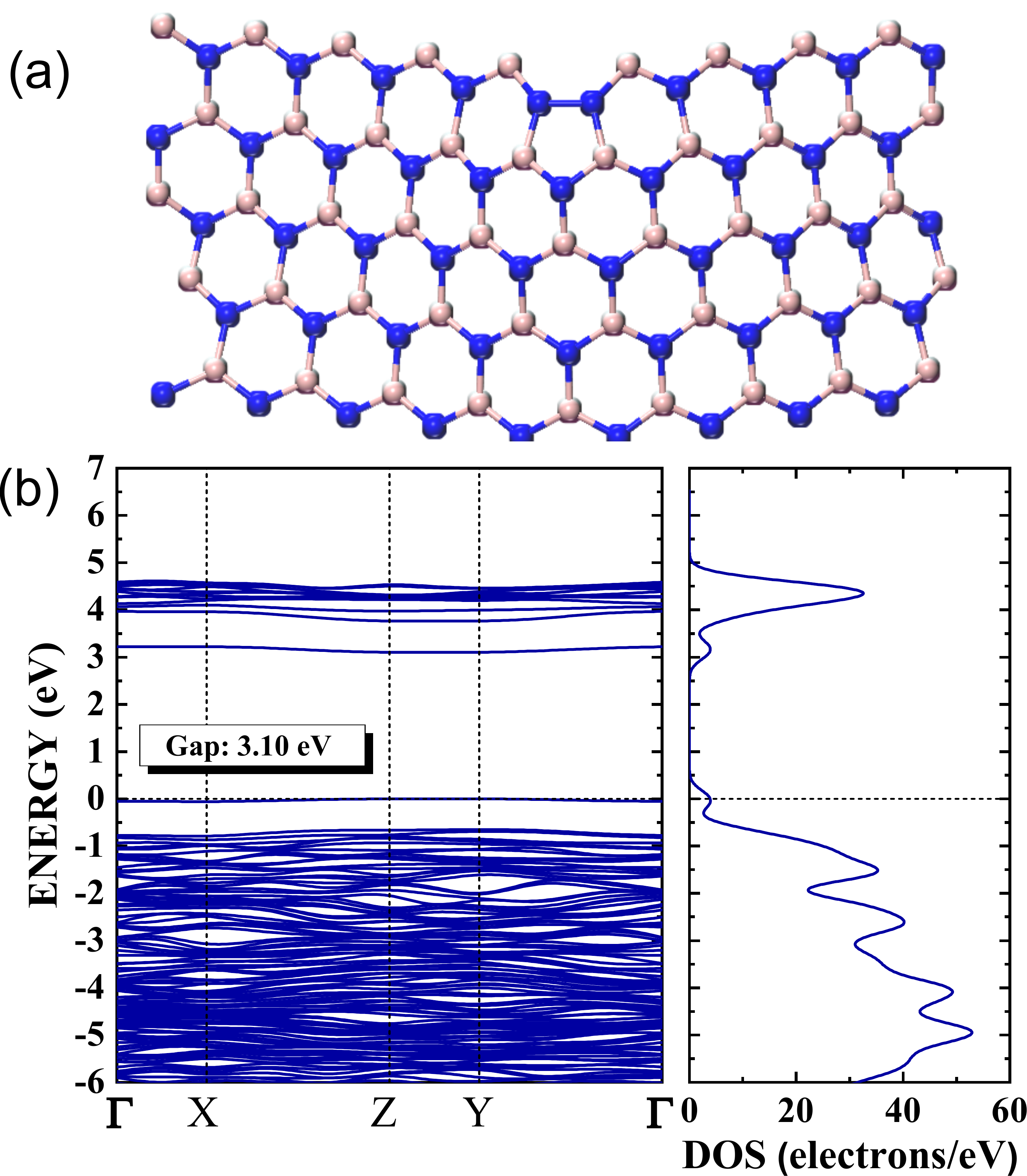}
	\caption{Representation of Kohn-Shan band structure and density of states of a structure with lower density of pentagonal-heptagonal pair of defects. (a) The unit cell of the structure. (b) The electronic structure revealing an energy gap of 3.10 eV.}
	\label{phabn2}
\end{figure}

Such conclusion is reinforced when we look to Fig. \ref{phabn2}, where we consider the band structure and the DoS of a system with a lower density of defects, which can be seen in the unit cell of the structure showed in Fig. \ref{phabn2} (a). We can see, in Fig. \ref{phabn2} (b), that the energy gap in this case increases to 3.10 eV. In the limit when the number of defects goes to zero, these states inside the energy gap should be absorbed in the valence and conduction bands and the insulator behaviour of the h-BN be recovered.

\subsection{Optical Properties}

\begin{figure}
	\centering
	\includegraphics[width=\linewidth]{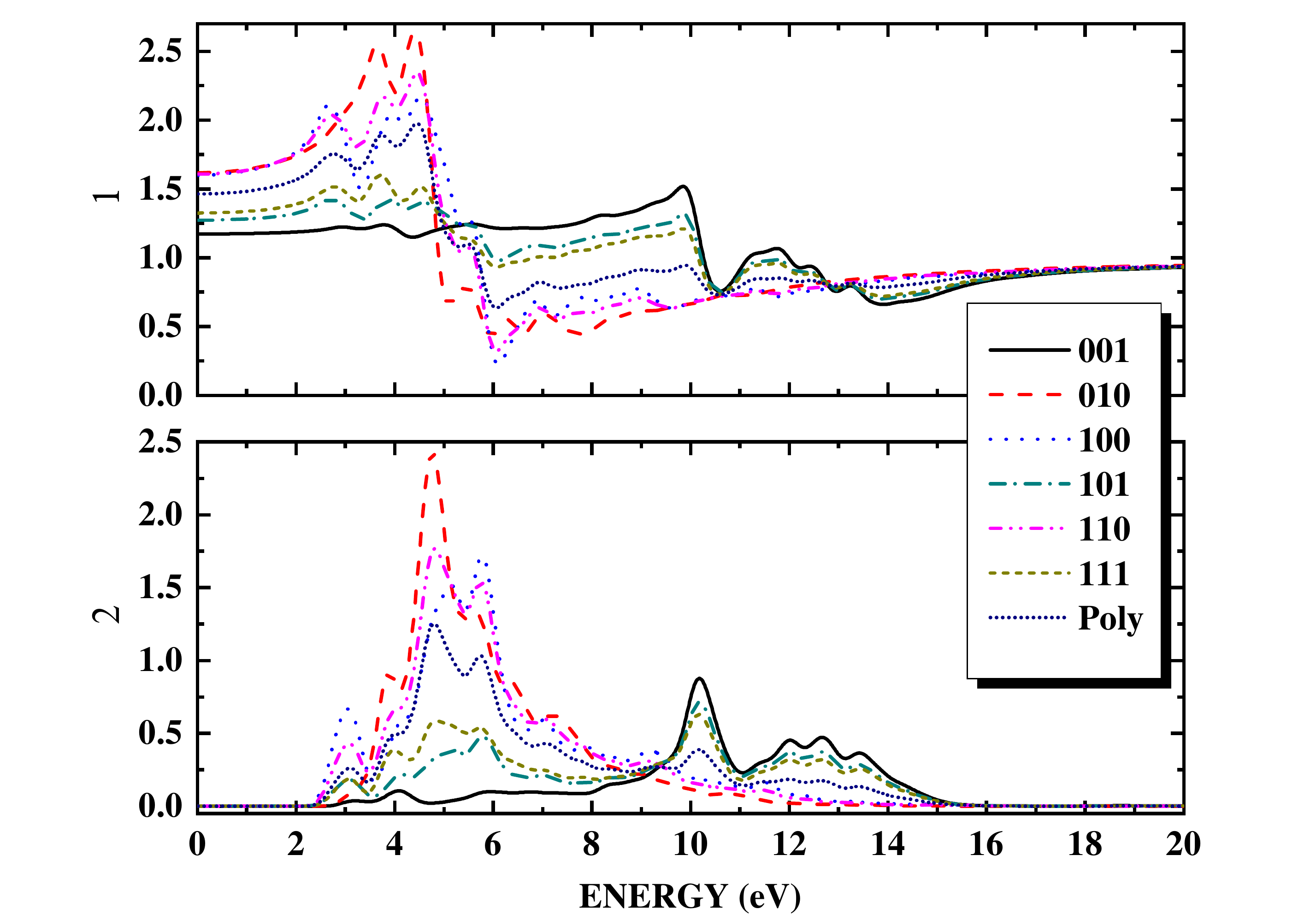}
	\caption{Real part ($\epsilon_1$) and imaginary part ($\epsilon_2$) of the complex dielectric function calculated with the GGA functional for the phaBN structure. The incidence of polarized light along the different crystalline and polycrystalline (poly) planes are shown.}
	\label{fig:die}
\end{figure}

Optical properties are essential for the characterization of a material, and this characterization is directly linked to the calculation of the complex dielectric function in terms of the energy (eV), where you have a real part  $\epsilon_1$ and an imaginary part $\epsilon_2$. In Fig. \ref{fig:die} we can observe how the dielectric function of the phaBN structure behaves. In the calculations we considered different directions of polarization and also the polycrystalline direction (Poly), and the exchange correlation functional adopted was GGA-PBE. It was possible to observe that the dielectric constant changes depending on the polarization direction chosen. The values of $\epsilon_0$ for the different directions were: $\epsilon_0$=1,6 for directions [010], [100] e [110]; 1,17 in the direction [001]; 1,27 in the direction [101]; 1,32 in the direction [111] and 1,46 for the Poly direction. It is important to mention here that we are identifying the crystalline directions using the Miller index, with the three numbers representing the three axis [$xyz$]. At this way, for instance, the [100] indicates a polarization in the $x$ direction, while [110] a polarization in a direction in the $xy$ plane that makes the same angle with both $x$ and $y$ axis.

There is a relation between the electronic transitions of the occupied and unoccupied states and the imaginary part $\epsilon_2(\omega)$ of the dielectric function, which is directly connected to the absorption of the material. From this effect the dielectric constant presents a causal response and we can thus map the behavior of $\epsilon_1(\omega)$ to $\epsilon_2(\omega)$ by means of the Kramres-Kroning transformations given by the following equations:
\begin{equation}
\epsilon_1(\omega) - 1 = \frac{2}{\pi}P\int_{0}^{\infty}\frac{\omega\;' \epsilon_2(\omega\;') }{\omega^2\;' - \omega^2} d\omega\;'
\end{equation}
and
\begin{equation}
\epsilon_2(\omega)= -\frac{2}{\pi}P\int_{0}^{\infty}\frac{\omega\;' \epsilon_1(\omega\;') }{\omega^2\;' - \omega^2} d\omega\;' .
\end{equation}
This dependence of $\epsilon_1(\omega)$ and $\epsilon_2(\omega)$ can be seen in Fig. \ref{fig:die}. The region of the peaks of the response functions are quite similar. For instance, for the polarization directions [010], [100] and [110], $\epsilon_2 (\omega)$ has more evident peaks around 5 eV due to the electronic transition of the B-2p valence states to the N-2p conduction states analyzed from the results of the partial density of boron and nitrogen atoms. 

\begin{figure}
	\centering
	\includegraphics[width=\linewidth]{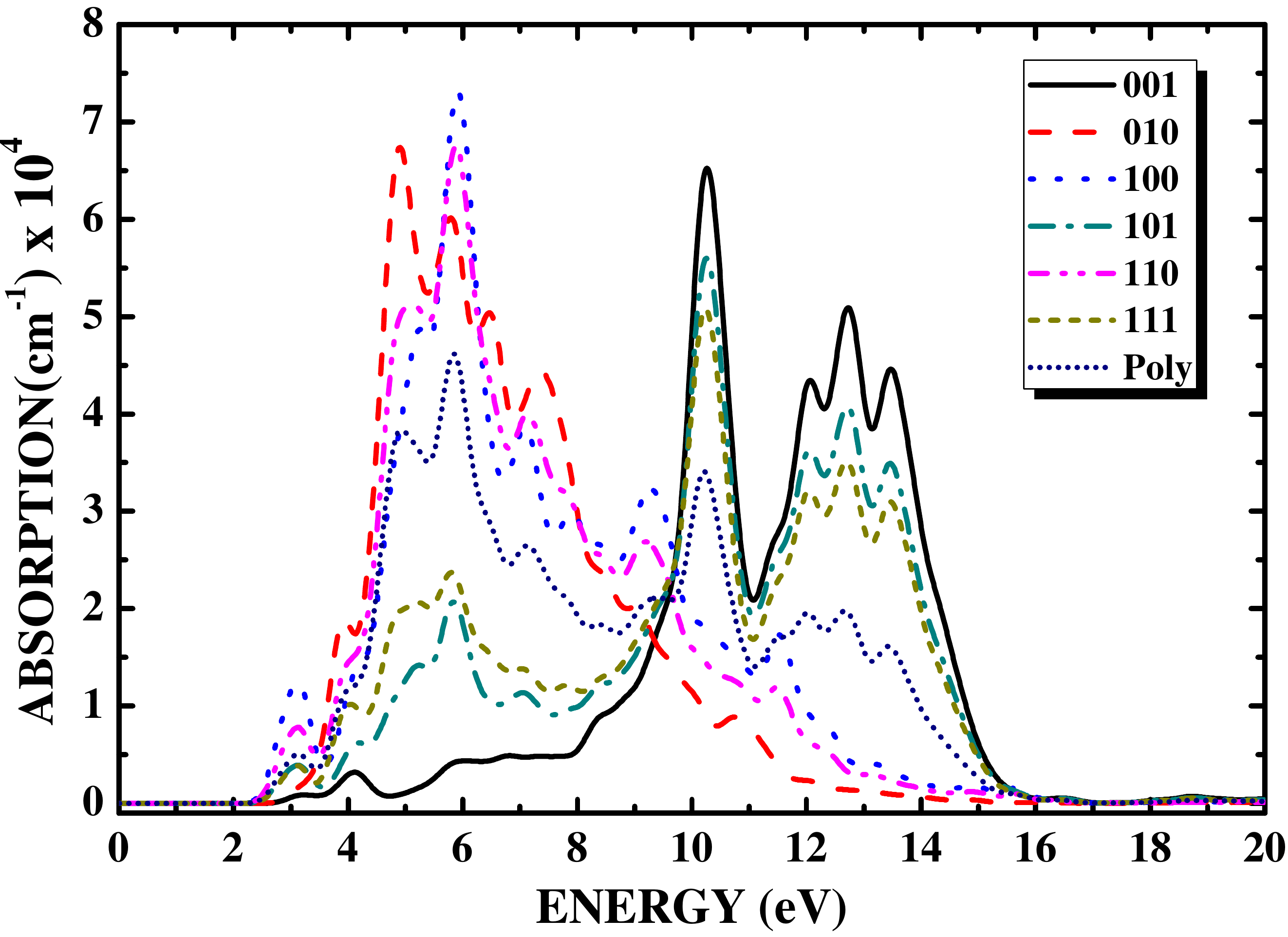}
	\caption{Optical absorption of the phaBN structure using the GGA-PBE exchange and correlation functional. The incidence of polarized light along the different crystalline and polycrystalline (poly) planes are shown. Absorption is maximum in the ultraviolet region of the electromagnetic spectrum.}
	\label{fig:absorcao}
\end{figure}

The absorption is represented in Fig. \ref{fig:absorcao}. We can note that for all direction of polarization the phaBN is transparent for visible and infrared light, while for ultraviolet light it exhibits strong optical absorption. For the polarization directions [010], [100], [110] and poly, the light is mostly absorbed in the UV-B and the smallest energies of the UV-C regions of the electromagnetic spectrum. On the other hand, the highest absorption peak for the [001], [101] and [111] directions of polarization is in the UV-C region ($\approx$ 10 eV). It means that the phaBN could be used, for instance, as a UV-B filter. 

\subsection{Vibrational and thermodynamic properties}

\begin{figure}
	\centering
	\includegraphics[width=\linewidth]{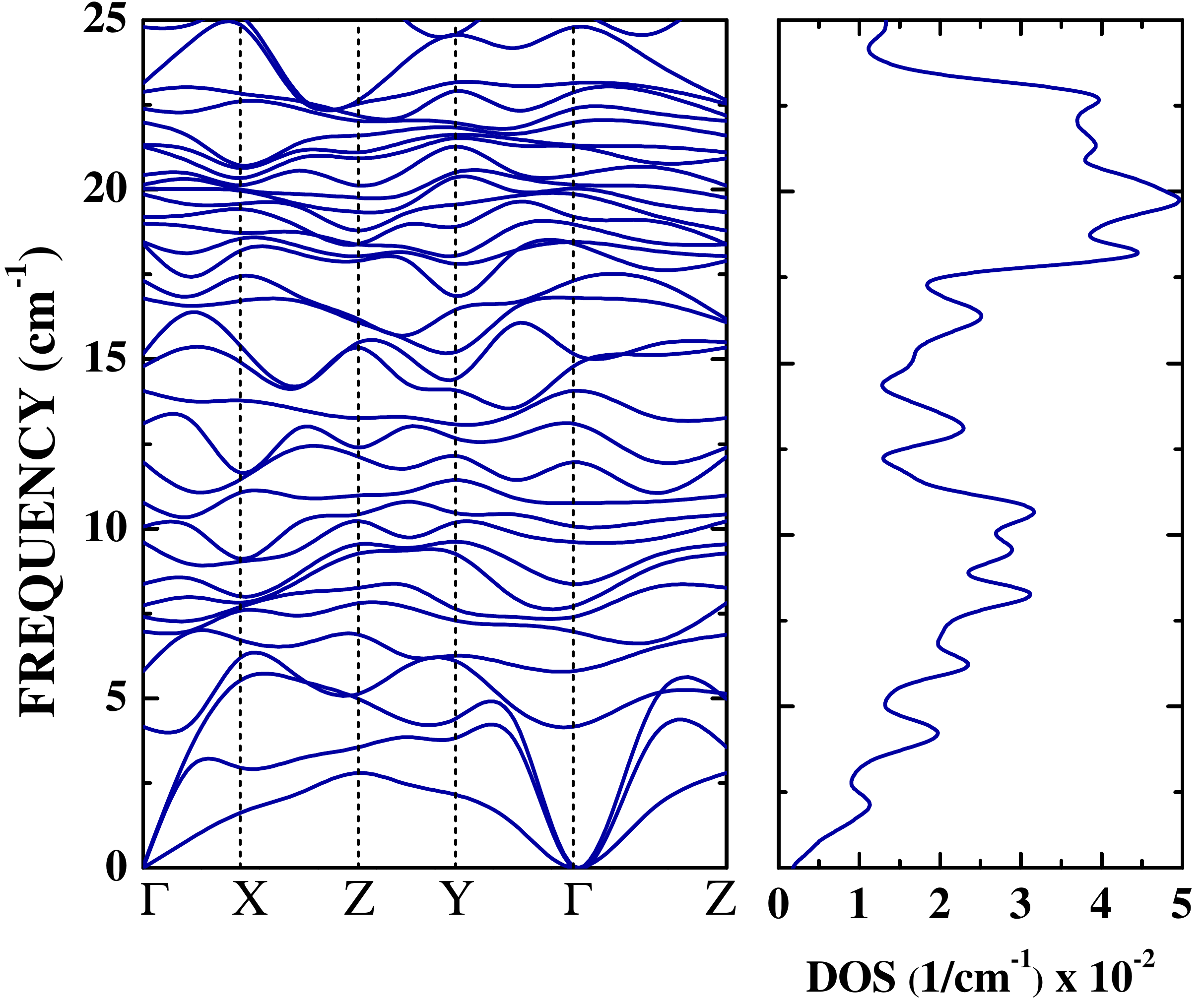}
	\caption{Phonon dispersion curves (left panel) and phonon density of states (right image) of the phaBN structure. We use the GGA-PBE exchange and correlation functional. We have a well-defined frequency spectrum in the range of 0 to 25 cm$^{-1}$, confirming the stability of the structure, since there are no negative frequencies in the first Brillouin zone.}
	\label{fig:fhonons}
\end{figure}

After the optimization process of the phaBN structure, we calculated the properties of phonons by means of the GGA-PBE exchange and correlation function, in order to see the stability of the structure. We used the conserved norm Pseudopotentials, which are required for calculations based on linear responses of phonon properties and polarizabilities. The energy convergence tolerance parameter for the force constants was defined as $10^{-4}$ eV /\AA$^2$; we also apply a non-analytical LO-TO correction for the dynamic matrix. On the left side of the Fig. \ref{fig:fhonons} we have the representation of the phonon dispersion curves along the points of high symmetry in the first Brillouin zone, where we consider a well-defined frequency range of 0 to 25 cm$^{-1}$. The contribution of three acoustic modes confirm the stability of the structure, since no imaginary frequencies were observed. We can also notice that there is no gap between the acoustic modes and the optical modes. These results are similar to the results of the h-BN phonon spectrum \cite{118,119}. On the right side of the Fig. \ref{fig:fhonons} we have the density of phonon states for the phaBN structure, which shows us the relation of the density peaks with the corresponding modes for each region. We can see that in the region between 15 and 25 cm$^{-1}$ is the region with the highest density of optical modes.

\begin{figure}
	\centering
	\includegraphics[width=\linewidth]{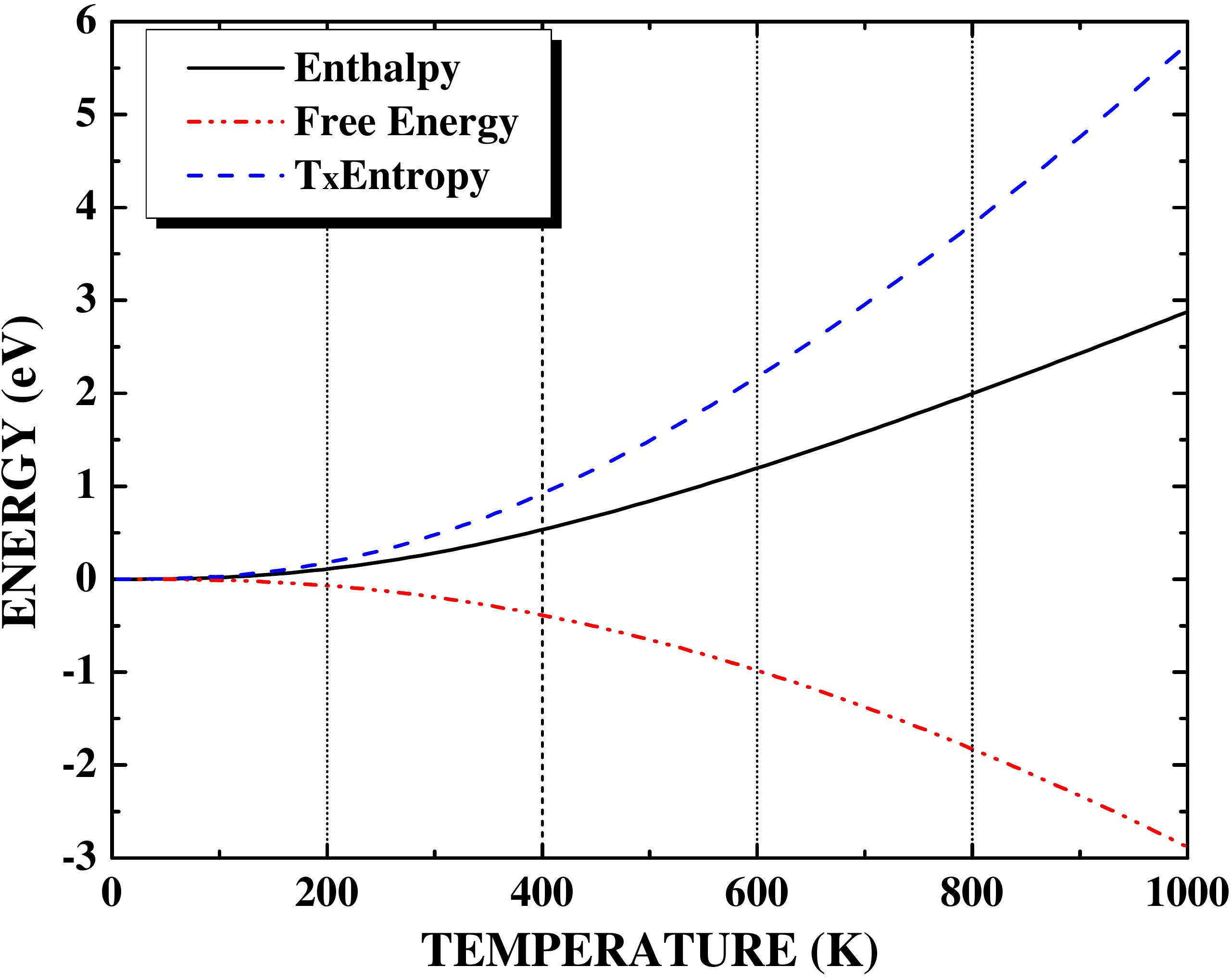}
	\caption{Representation of thermodynamic potentials given by: enthalpy (solid black line), free energy (red dotted line), temperature times entropy in units of eV (dashed blue line), calculated as a function of temperature for the structure of phaBN.}
	\label{fig:termodinamicas}
\end{figure}

\begin{figure}
	\centering
	\includegraphics[width=\linewidth]{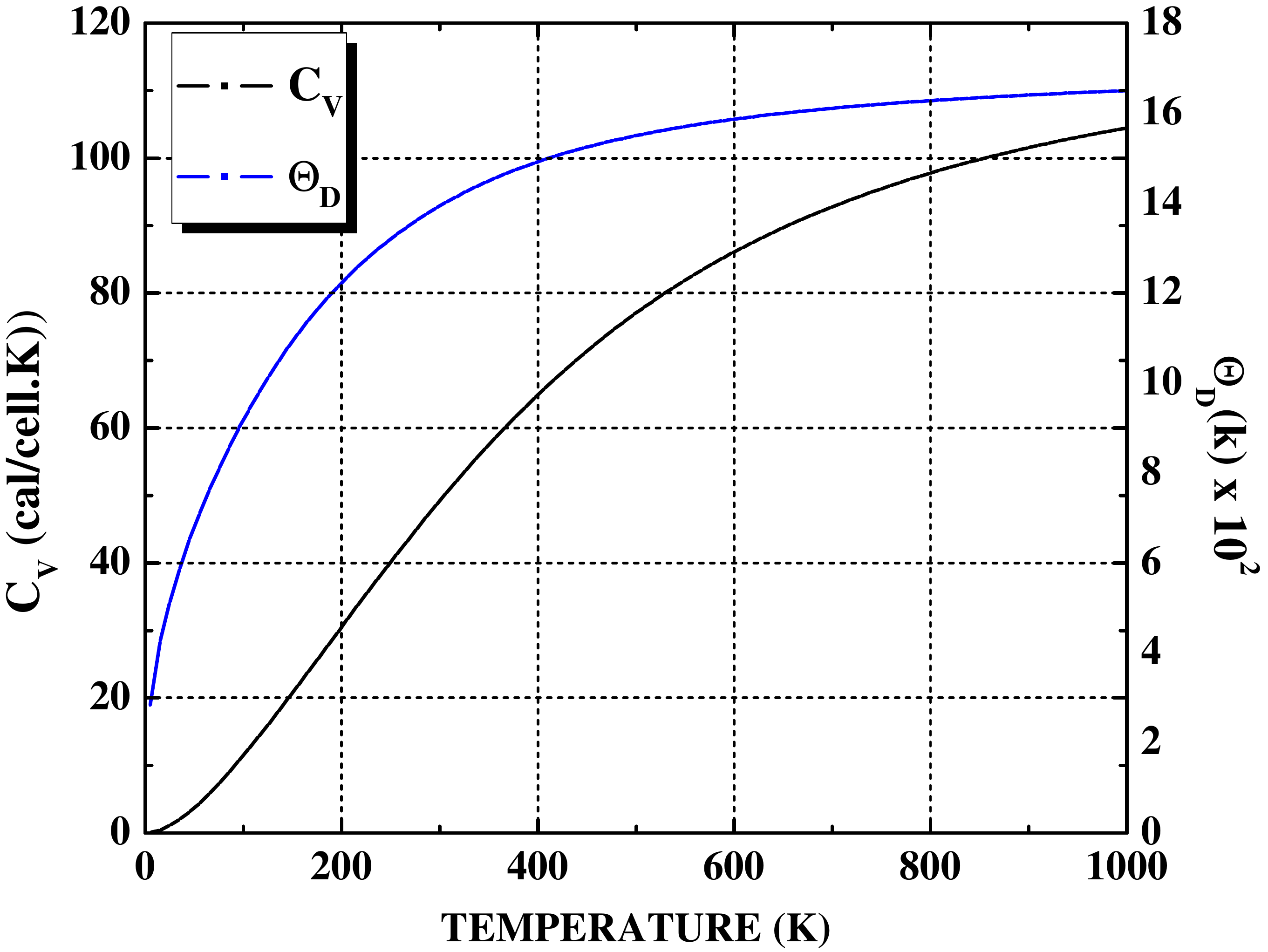}
	\caption{Heat capacity at constant volume $C_V$ as a function of temperature (solid black line). The blue line, using the right scale, represents the temperature dependence of Debye $\Theta_D(T)$.}
	\label{fig:cv}
\end{figure}

The thermal properties are shown in Figs. \ref{fig:termodinamicas} and \ref{fig:cv}. Thermal expansion was not considered because it does not play an important role in solid materials. What is noticeable from Fig. \ref{fig:termodinamicas} is that the enthalpy (dashed black line) grows with temperature and exhibits almost linear behavior. The temperature times the entropy (solid blue line) also grows as the temperature increases, indicating that the structure is tending to a state of maximum equilibrium. The Gibbs free energy (dotted red line) is decreasing as the temperature increases, which means that the process of forming the structure is spontaneous. 

We have in the Fig. \ref{fig:cv} the representation of the heat capacity at constant volume $C_V$ (solid black line) and the behavior of the temperature variation of Debye $\Theta_D$ (line Blue trace) as a function of the absolute temperature. It is well known that the heat capacity increases as the temperature increases to its maximum around the Dulong-Petit limit. On the other hand, $\Theta_D$ also increases as the temperature increases, having a maximum around 1700 K.

\section{Conclusion}

In summary, we performed first-principle calculations in a density functional theory framework to investigate the electronic, optical, thermal and vibrational properties of the phaBN structure. The phaBN presents an energy gap of 2,739 eV, which is almost half of the energy gap of h-BN, which means that it is a semiconductor. Such reduction in the energy gap occurs due to the presence of new electronic states inside of the energy gap, which are induced by the pentagonal and heptagonal defects in the structure.  The complex dielectric function presents variations in its real and imaginary part, thus being anisotropic in relation to the direction of light. 
The optical absorption revealed that the phaBN is transparent for visible and infrared light and that it has a maximum absorption in the UV region of the spectrum, which means that phaBN could be used as a UV filter. 
The spectrum of phonons presented a well-defined frequency region, showing that there are no negative frequencies throughout the first Brillouin zone, indicating the stability of the system. Finally, the thermodynamic properties show us that the formation process of the structure is spontaneous.

{\bf Acknowledgments}: This work was partially supported by Capes, CNPq and Alexander von Humboldt Foundation.

The raw/processed data required to reproduce these findings cannot be shared at this time due to technical or time limitations

\end{document}